\documentclass[prd,eqsecnum,noshowpacs,nofootinbib,notitlepage]{revtex4-1}
\usepackage{amsmath,amssymb,amsthm}
\usepackage{amsfonts}
\usepackage{comment}
\usepackage{color}
\usepackage{graphicx}
\usepackage{mathptmx}


\newtheorem{theorem}{Theorem}
\newtheorem{definition}{Definition}

\newtheorem{corollary}{Corollary}

\newtheorem{proposition}{Proposition}
\newtheorem{lemma}{Lemma}

\begin{document}
\title{ Black hole shadow and Wandering null geodesics}

\author{Masaru Siino}
\email{msiino@th.phys.titech.ac.jp}
\affiliation{Department of Physics, Tokyo Institute of Technology, Tokyo 152-8551, Japan}

 \begin{abstract}
The role of the wandering null geodesic is studied in a black hole spacetime.
Based on the continuity of the solution of the geodesic equation, the wandering null geodesics commonly exist and explain the typical phenomena of the optical observation of event horizons.
Moreover, a new concept of `black room' is investigated to relate the wandering null geodesic to the black hole shadow more closely.
 \end{abstract}

\maketitle
\section{Introduction}
The black hole shadow\cite{Falcke:1999pj,Broderick:2009ph,Broderick:2010kx} observed as the real image of photons around the massive black hole at the center of M87 was excitingly reported by the Event Horizon Telescope (EHT) -- a planet scale array of eight ground based radio telescopes forged through international collaboration \cite{Akiyama:2019cqa,Akiyama:2019brx,Akiyama:2019sww,Akiyama:2019bqs,Akiyama:2019fyp,Akiyama:2019eap}.
Nevertheless, of course, by definition of the black hole we cannot see its event horizon directly.
Black hole region is defined as the complement of the causal past of future null infinity, and the event horizon is defined as the boundary of the black hole region. 
Then any causal past set of us, which is the observer on the future null infinity does not contain the event horizon. There is not any direction among our line of sights corresponding to the event horizon. Therefore, we should be careful to understand the meaning of the image of the black hole shadow.

On the other hand, when we consider physically realistic black hole, which is not the rigorous Schwarzschild spacetime, the most significant fact we should concern about the black hole, would be that the black hole is not static.
Of course, the simplest essential picture of the shadow is interpreted by the appearance of a photon sphere\cite{photonsphere,Hod2013,Hod2018}.
In Schwarzschild spacetime, the unstable circular orbits of photons form a bright photon sphere. And then, the absence of the orbit in which light ray goes into the photon sphere and succeedingly goes out from it, implies the centered dark region\cite{Sanchez1978,Decanini2010}.
Nevertheless, in general realistic black hole spacetime, the concept of the photon sphere fails and the formation of the black hole will become significant.

By monitoring the morphology of the event horizon, it was suggested\cite{Wielgus2020} that the shadow of a black hole does not show any dynamical nature. 
We are still expecting, however, that there still remain any remarkable dynamical aspects of the black hole shadow which we can observe during our real life time.
The expectable dynamics that will be oscillation, movement, formation, merging, and so on.
For example, though the oscillation may be treated perturbatively, we are not hopeful to understand them completely without knowing any general extended treatment of the black hole shadow beyond the Schwarzschild spacetime.

On the other hand, the concept of the photon sphere is generalized by the wandering set into the general black hole\cite{Siino2020}.
Then it is suggested that the wandering null geodesic, which is a complete null geodesic possessing an infinite number of conjugate points, will be accompanied by the accumulating null geodesics coming from the photon sources at here and there, under the existence of a lot of conjugate points.
We are wondering what is the condition for the appearance of the wandering null geodesic. Since the key concept will be the conjugate point, we should discuss the effect of the curvature to null geodesic congruence. 
In the Schwarzschild spacetime, however, the fact is that any spacetime point of the outside of the event horizon has such a wandering null geodesic.
That means the optical signal from all of the spacetime can be included by the bright region of the black hole shadow. Is it also true for general black holes?

One of the purposes for the present article is to investigate such a diffusional nature that the wandering null geodesics come from here and there. Especially, we would like to discuss its observational meaning for the black hole shadow.
In the next section, we will remind the definition and the fundamental concept of the wandering null geodesic and wandering set as a generalization of the photon sphere.
In the third section, we demonstrate the commonplaceness that the wandering null geodesic can be starting from here and there in the black hole spacetime, under the global hyperbolicity.
And then, we discuss the meaning of the black hole shadow observation under the commonplaceness in the fourth section. Moreover, we will attempt to relate the wandering null geodesic closely to black hole shadow introducing the concept of black room.
The final section is devoted to summary.

Strictly speaking, the meaning of black hole shadow should be the dark region of the black hole image. Therefore, it does rigorously not include the bright sphere.
Nevertheless, this distinction is not so productive.
In the present article, we expect that the reader regards the word black hole shadow as including both dark region and bright region of the black hole image.

\section{definition of wandering null geodesic}
To generalize the photon sphere, one may attempt to consider a photon lying on a timelike hypersurface\cite{Schneider2018,Cao:2019vlu,Mishra2019,Bisnovatyi-Kogan2018, Mars2017}. 
In the Schwarzschild spacetime, the characteristic of the null geodesic equation was reduced to a single potential function. So, one would expect the similar analysis gives a light to shine on a road to generalized photon sphere.

There, the timelike hypersurface could divide the spacetime into two regions where gravity is strong and weak\cite{Yoshino2020}. In the first place, however, it is not clear even whether the generalization of the photon sphere should be a timelike hypersurface whose spatial section is any closed surface (naively, we may expect it to be a sphere) or not, in general situations.
At the present, we think those are doubtful since the essence of such an orbital kinematics will rely on not only the gravitation in radial direction, but also in angular direction which will deform the orbit evolutionally, in general.
Therefore, we rather give attention to a single null geodesic than that dividing surface, firstly.


Since the purpose of the generalization of the photon sphere is to explain how the black hole shadow is structured, we will think the generalization must not depend on the spacetime coordinate, especially in the choice of the spatial hypersurface. According to the general covariance, timelike directions for each coordinate system are on an equal adequateness. So, it is difficult to determine spacelike hypersurface without any spacetime symmetry or conservation law.

Even if we concentrate on a single null geodesic, how can we judge the null geodesic is on an orbit or not without a concept of any special time coordinate?
Consequently, we will consider that the concept of the generalized photon sphere should be far away from the concept of coordinates and the most natural substitution of it would be a concept related to causality. 
Indeed, the definition of a black hole is given as the complement of the causal past of the future null infinity.
Then also the black hole shadow as the fundamental nature of the black hole is expected to be discussed as the causal nature of the spacetime.

Though one of physical key points of the black hole shadow in static spherically symmetric spacetime was that many null geodesics are accumulated to a single unstable circular photon orbit\cite{Virbhadra2000}, at the first onset we will anyway concentrate on a causal nature of a single null geodesic which can substitute the circular orbit.

Considering an asymptotically flat spacetime~\cite{Hawking:1973uf,Wald:1984rg}, inextendible complete curves go toward   the boundary of the spacetime manifold composed of future (past) null infinity $\cal I^+,\cal I^-$, future (past) timelike infinity $i^+, i^-$ and spatial infinity $i^0$.
The simplest causal structure will be the asymptotically simple where every null geodesic will terminate only on the future null infinity.
Nevertheless, as well known the black hole spacetime is not the case since the null geodesic going into the event horizon never can escape from it.

The remarkable fact is that there are null geodesics which go toward future timelike infinity $i^+$, which is the stable circular orbit $r=3m$ in Schwarzschild spacetime.
Though such a geodesic will not be allowed in asymptotically simple spacetimes~\cite{penrose1986spinors}, general black hole spacetimes may have such null geodesics under the condition of weak asymptotic simpleness.
And then the accumulation of photon orbits will be discussed forcusing on whether there are many parallel null geodesics which go to escape to the future null infinity (or coming from the past null infinity).

 In the sense of conformal completion, however, the unphysical manifold $\overline{M}$ does not include the point of timelike infinity $i^+$, since the conformal boundary is not smooth there.
To include $i^+$ in the boundary of the manifold, one may consider c-boundary~\cite{Geroch:1972un}.
 Then we see the boundary will have Hausdorff topology, but it is not easy to handle it.
Therefore, it would not be a good choice to define the concept that geodesics go to $i^+$.

In spite of that, we will consider null geodesics not falling into a black hole and not escaping to future null infinity.
We may call such a null geodesic a `neutral' null geodesic $\gamma_n$ as a generalization of the unstable circular photon orbits.
To examine such a neutral null geodesic, we will analyze the null geodesic congruence\cite{Wald:1984rg}. Then we will soon be aware that the neutral null geodesic will be accompanied by the infinite number of conjugate points, if the null geodesic is complete and inextendible\footnote{In the present work, we only consider inextendible one since the parameter of the null geodesic for light ray is to be set to affine parameter.}\cite{Siino2020}.
And then, we will define a wandering null geodesic as the generalization of the unstable circular orbit for photons. We consider a future (past) complete null geodesic with repetitive infinite number of conjugate points starting from $p$ to the future (past) direction. Then it is named a future (past) wandering null geodesic from $p$.


As we are considering a globally hyperbolic spacetime, the existence of conjugate points implies that such a null geodesic come into the inside of the chronological future of the starting point.
Then the same logic suggests that along this null geodesic conjugate points repetitively appear. 
This situation can be easily illustrated in the case of Schwarzschild spacetime (see Fig. 2 in reference\cite{Siino2020}).

The concept of the conjugate point is a natural result, because under the global hyperbolicity the following theorem is well known\cite{Hawking:1973uf}\cite{Wald:1984rg}.

 
\begin{theorem}
Any point on the boundary of a causal future $J^+(p)$ can be connected to $p$ by the null geodesic without conjugate point.
\label{thm:1}
\end{theorem}

This theorem means, in a sense, the formation of conjugate points implies that the null geodesic enters the chronological future of the starting point under the global hyperbolicity (see Fig. 3 in reference \cite{Siino2020}). 
This situation is of small variation for the null geodesic generator of the chronological future\cite{Hawking:1973uf}\cite{Wald:1984rg}.
Roughly speaking, the existence of conjugate points suggests that the null geodesics is passing round and round around the black hole, repetitively visiting the conjugate points. \footnote{Of course, one may argue that the conjugate point is not necessarily required for the null geodesic to enter  the chronological future of the starting point (topological non-triviality may globally cause it, for example in a static spacetime of flat $S_1\times R^2$ spatial section).}
Then we think that the usual wandering of the null geodesic around the black hole region is related to the conjugate point of complete null geodesics.

As discussed in \cite{Siino2020}, the correct correspondence of the circular orbit is not general wandering null geodesics in rigorous sense.
A totally wandering null geodesic is defined as a future past complete null geodesic possessing an infinite number of conjugate points both in future and past directions.
For the generalization of the photon sphere, we consider a wandering set, which is the set of the totally wandering null geodesics.
Furthermore, one may manipulate the null geodesics in order to investigate various aspects of the wandering nature.
For example, for the sake of the determination of the formation of the wandering null geodesics, we will truncate the segment of the future wandering null geodesics at the first or the second conjugate point.
According to the definition of the wandering null geodesic, in Schwarzschild spacetime, we see that there are starting points of the wandering null geodesic on all over the world commonly, as it can be easily seen by the spherical symmetry and the static nature.

\section{commonplaceness of wandering null geodesic}

 Let us focus on an asymptotically flat black hole spacetime.
Through the present work, to assure the causality we impose global hyperbolicity on the spacetime or a part of the spacetime which contains whole of the outer region of the black hole\footnote{c.f., cosmic censorship conjecture}. Such a part of the spacetime is called a strongly asymptotically predictable spacetime, and we assign the character $M$ to it \cite{penrose1986spinors}. Note that more precisely, an asymptotically flat spacetime $(M,g)$ is a strongly asymptotically predictable spacetime when  there is sufficiently large open globally hyperbolic region $V$ in $M$ containing $J^-({\cal I}^+)$.
  As a consequence of asymptotic flatness, the unphysical spacetime manifold $\overline{M} $ possesses spatial infinity $i^0$ and future null infinity ${\cal I}^+$, which is defined as the causal future of $i^0$.  
Moreover, future null infinity ${\cal I}^+$ admits a null coordinate $u$ as ${\cal I}^+ \simeq \{(0,\infty)\}\times S^2\sim \{(u, x^1, x^2)|u\in (0,\infty)\}$.
 
 From the global hyperbolicity, there is a family of Cauchy surfaces ${\cal C}[t]$ as a family of time slices of the spacetime such that $M\simeq {\cal C}[\cdot]\times [ 0,\infty )$.  Then the family of time slices is regarded as complete, since $\overline{M} \supset J^-({\cal I}^+)$ even if there is a timelike or null singularity inside the event horizon.
For an initial Cauchy surface ${\cal C}[0]$, we assume that there already exists the black hole region: ${\cal C}[0]\cap (M\backslash J^-({\cal I}^+))\neq \emptyset$, which should finally settle to a single spatially connected region. We also assume nonevaporativity of black holes \footnote{Even if the spacetime is globally hyperbolic and asymptotically flat, this assumption is violated by the thunderbolt singularities\cite{Hawking1993}.} that is, for an arbitrary point $p $ on $ {\cal C}[0] $, there exists a causal curve from $p$ to the interior of the black hole region $B:=M \backslash \overline{J^-({\cal I}^+)}$.

Now we discuss the set of continuous future directed causal curves starting from $p$ and ending at $q$, $C(p,q)$.
By the global hyperbolicity, $C(p,q)$ is compact~\cite{Geroch:1970uw} with respect to the topology generated by the following basis, 
 \begin{align}
 O(U) = \{ \lambda \in C(p,q) | \lambda \subset U\},\label{top},
 \end{align}
where $U$ is an arbitrary open subset of the spacetime manifold $M$.
Here it should be noted that $C(p,q)=\emptyset$ for $q\not\in J^+(p)$.
Succeedingly we determine another set of curves from $C(p,q)$ for a Cauchy surface $\Sigma$ by
\begin{align}
 C(p,\Sigma)=\bigcup_{q\in \Sigma}C(p,q),
\end{align}
where we define the topology of $C(p,\Sigma)$ replacing the point $p$ with a Cauchy surface $\Sigma$ in (\ref{top}). One can show that $C(p, \Sigma)$ is compact.

Here we define the subset of $C(p,\Sigma)$, which plays an important role in the proof of commonplaceness of wandering null geodesics:
 \begin{definition}
    $N(p,\Sigma)$ is the set of the null geodesics starting from $p$ to $q\in \Sigma$.
 \end{definition}
 Here it should be noticed that while any causal curve connecting $p$ to $q$ must be a null geodesic if $q\in J^+(p)\backslash I^+(p)$ (theorem \ref{thm:1}), $N(p,\Sigma)$ is composed of not only such null geodesics since it can contain a null geodesic with a conjugate point of $p$.
 
In the following, we investigate properties of $N(p,A)$ in analysis of ordinary differential equation (ODE).
Here we had better be careful about the smoothness and continuity of the spacetime, since we might consider various spacetimes, e.g., at the surface of a star the matter field might not be continuous or Kastor-Trascen spacetime is not so smooth\cite{Nakao1995}.
Now we will demonstrate that the homotopical nature between the initial value and its evolution in a neighborhood of a unique solution.
With the existence and uniqueness of the solution for the ODE, we quote the following known theorem which is essential for existence of wandering null geodesics.
\begin{theorem}[continuity of ODE \cite{Coddington1955}]
Let $\tau_0, \tau_1(\tau_0<\tau_1)$ be real fixed numbers, and $(\tau_0,\xi_0)$ a fixed point in the $(n+1)$-dimensional $(t,x)$ space. Denote by $U_0$ the set of all points $P_0:(\tau_0,\xi)$ such that 
\begin{align}
|\xi-\xi_0|<b_0, \ \ \ (b_0>0).
\end{align}
Suppose that through each point $(t,x)$ in the region
\[ V:\ \ \ \tau_0\leq t\leq \tau_1\ \ \ |x-\xi_0|<b \ \ \ (0<b_0\leq b) \]
there exists a unique solution of 
\begin{align}
\frac{dx}{dt}=f(t,x),
\label{eqn:ODE}
\end{align}
$f$ being continuous on $V$. Let $\varphi=\varphi(t,\tau_0,\xi)$ be the solution of (\ref{eqn:ODE}) passing through $P_0\in U_0$. 
Let $b$ be sufficiently large so that $(t,\varphi(t,\tau_0,\xi))\in V$for$ |\xi-\xi_0|<b_0,\ \tau_0\leq t\leq \tau_1$. 
Let $U_1$ denote the set of all points $P_1:(\tau_1,\varphi(\tau_1,\tau_0,\xi))$, where$(\tau_0,\xi)\in U_0$. 
Then the mapping $T$ which assigns the point $P_1$ to each point $P_0\in U_0$ is a homeomorphism of $U_0$ onto $U_1$.
\label{thm:ODE}
\end{theorem}
\vspace{0pt}


Now we apply Theorem \ref{thm:ODE} to $N(p,\Sigma)$ and discuss the existence and uniqueness of the solution of the ODE.
Incidentally, the existence of the solution for the ODE is locally demonstrated by Cauchy and Peano, and we should impose the value of $f$ to be bounded as $|f|<\cal M$ to extend the solution globally.
Nevertheless, by the global hyperbolicity the existence of the solution is assured since $C(p,\Sigma)$ including $N(p,\Sigma)$ is compact so that the Cauchy series of a curve $\lambda_n\in N(p,\Sigma)$ is convergent.
As well known, for the uniqueness, the Lipschitz continuity will be adopted as a sufficient condition. Of course the necessary sufficient condition for the uniqueness of the solution for the ODE is also given by Okamura\cite{OkamuraUnique}.

We define $S[t]=\{\lambda\cap{\cal C}[t]|\lambda\in N(p,{\cal C}[t])\}$ and should be careful about the fact that $T$ in theorem \ref{thm:ODE} is not one to one in globally, since there might be conjugate points.
Thus a direct consequence of global hyperbolicity, one can show that  $S[t]$ is arc-wise connected.
 \begin{corollary}
 Suppose $M$ is globally hyperbolic spacetime, where the geodesic equation satisfies the condition for the theorem \ref{thm:ODE}.  Then $N(p,{\cal C}[t])$ is also arc-wise connected in a topology of eq.(\ref{top})
  $S[t]$ is arc-wise connected.
  \label{prop:AC}
 \end{corollary}
 {\bf Proof:}
From the theorem \ref{thm:ODE}, there is a topological immersion of a sphere $\Phi[t]:S^2\mapsto {\cal C}[t]$. 
 Then $S[t]$ is arc-wise connected, and $N(p,{\cal C}[t])$ is also arc-wise connected in a topology of eq.(\ref{top}).
\qed\\

\begin{figure}[hbtp]
  \begin{center}
\includegraphics[height=10cm]{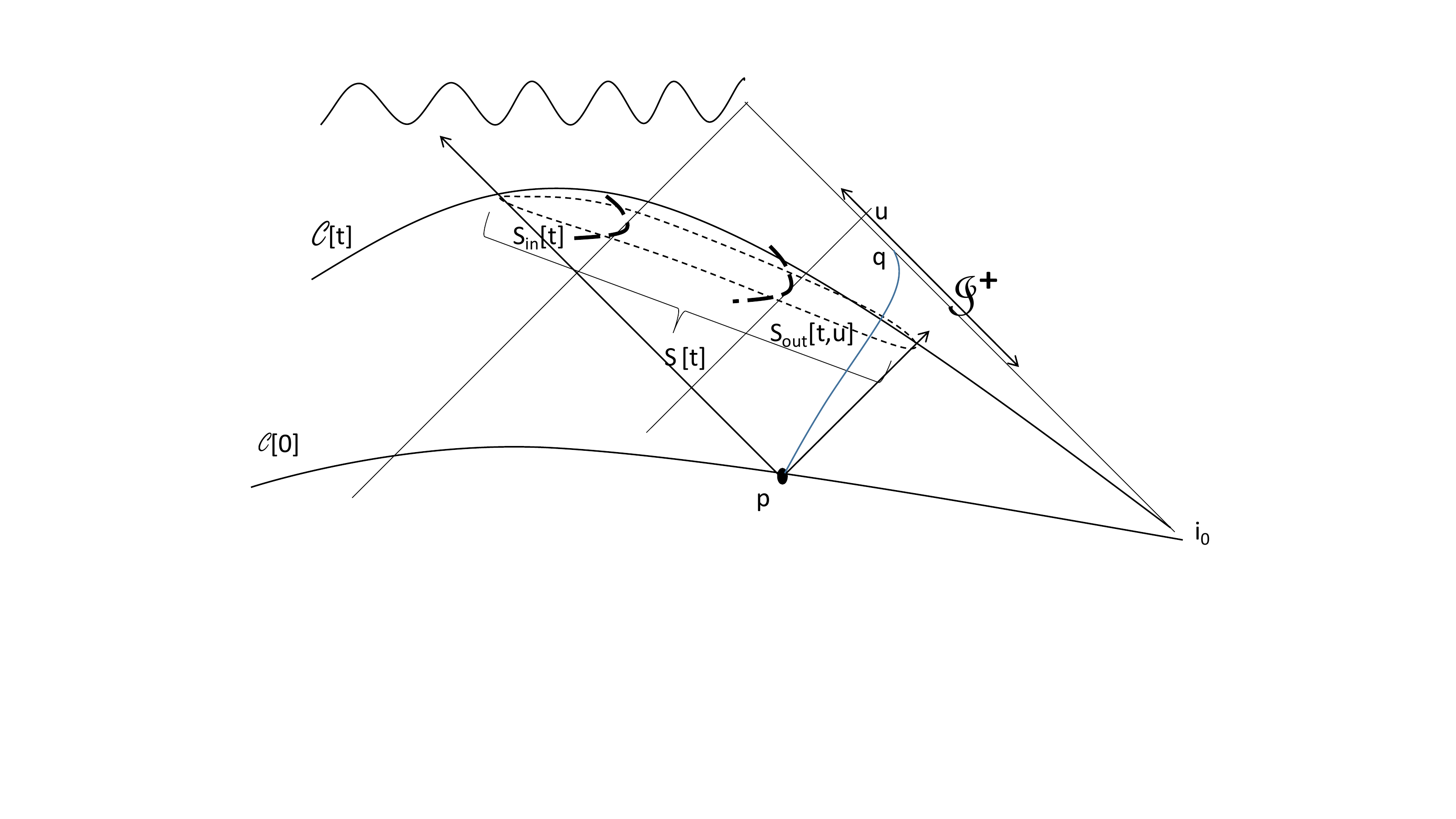}
  \end{center}
 \caption{The conformal diagram of a part of an asymptotically flat black hole spacetime: $S_{in}[t], S_{out}[t,u]$ on ${\cal C}[t]$ are drawn. 
} 
   \label{fig:subset}
 \end{figure}

Thus we are ready to prove the theorem of commonplaceness for a wandering null geodesic. Firstly, we consider the neutral null geodesics.
To prove the existence of the neutral null geodesic, we define a subset ${\cal I}^+[u_0]=\{(u,x^1,x^2)\in {\cal I}^+|u<u_0 \}\subset {\cal I^+}$ and following close subsets of $S[t]=\{\lambda\cap{\cal C}[t]\ |\ \lambda\in N(p,{\cal C}[t])\}$:
\begin{definition}
  \begin{align}
&S_{out}[t,u]=S[t]\cap J^-({\cal I}^+[u]),\\
   &S_{in}[t]=S[t]\cap \overline{B}=S[t]\cap \overline{(M \backslash  J^-({\cal I}^+))}.      
  \end{align}
  \label{def:subset}
  \end{definition}
Then an element of $N(p, S_{out}[t,u])$ represents a null geodesic going toward future null infinity and that of $N(p, S_{in}[t])$ represents a null geodesic falling into a black hole.
Our purpose is to demonstrate the existence of a null geodesic which is included by neither $N(p, S_{out}[t,u])$ nor $N(p, S_{in}[t])$.
The situation is illustrated in the Fig. \ref{fig:subset}.

 We can show that $S_{in}[t]$ and $S_{out}[t,u]$ is not empty for sufficiently large $u$ and $t$, and they are not intersecting.
 \begin{proposition}
  For a point $p$ on ${\cal C}[0]$ and the outside of the black hole region, there exist $t_0$ and $u_0$ such that $S_{out}[t,u]$ and $S_{in}[t]$ are not empty for any $t>t_0$ and $u > u_0$. Moreover $S_{out}[t,u]\cap S_{in}$ is empty.
  \label{prop}
 \end{proposition}
{\bf Proof:}
Since $p$ is a point in $J^{-}({\cal I}^+)$, for any $q$ on ${\cal I}^+$, there is a causal curve connecting $p$ and $q$. Let $u_0$ be a $u$-coordinate value of the point $q$. Then $S_{out}[t,u]$ is not empty for any $u > u_0$.

Since the initial time slicing ${\cal C}[0]$ is achronal, $B \cap {\cal C}[t]$ is not a subset of $J^+(p)$ for arbitrary $t>0$, where $\partial J^+(p)$ is generated by the null geodesics starting from $p$.
From the assumption of the nonevaporativity, there exists $t_1$ such that $(B \cap {\cal C}[t])\cap J^+(p)$ is not empty for arbitrary $t>t_1$.
Since we consider the spacetime which `settles down' to a single black hole state, there exists $t_2$ such that $B \cap {\cal C}[t]$ is arc wise connected for arbitrary $t>t_2$.
From these three facts, there exists $t_0$ such that $(\overline{B} \cap {\cal C}[t])\cap S[t] =S_{in}[t] $ is not empty for any $t>t_0$.

If $S_{in}\cap S_{out}$ is not empty, consider $r\in S_{in}[t]\cap S_{out}[t,u]$. From $r\in S_{out}[t,u]$,  there is a $q'\in \cal I^+$ such that $r$ and $q'$ are connected by a null geodesic. Then in a neighborhood ${\cal U}_q'$ of $q'$ on $\cal I^+$, there is $q''\in {\cal U}_q'$ such that $r$ and $q''$ is connected by a timelike curve.
Therefore $r$ is in an interior of $J^-({\cal I}^+)$. 
From $r\in S_{in}[t]$, $r$ is in $B=M\backslash J^-({\cal I}^+)$ or $\partial B$.
Since they are inconsistent, $S_{in}\cap S_{out}[t,u] =\emptyset$.
 \qed\\

Then we claim that there exists a null geodesic which intersects neither $S_{out}[t,u]$ nor $S_{in}[t]$ for arbitrary large $u$ and $t$. 
Now we are able to give the concrete definitions of such a neutral null geodesic to match the theorem:

\begin{definition}[neutral null geodesic]
 A null geodesic which starts from the point $p$ is called a {\it neutral null geodesic} starting from $p$
  if for any $u$ there exists a sufficiently large $t$ such that it does not cross the event horizon and ${\cal C}[t]\cap J^-({\cal I}^+[u]) $.
\end{definition}

Then we have a following lemma,
\begin{lemma}
For any $p$ on ${\cal C}[0]$ outside of black hole region, there exists a neutral null geodesic starting from p.
 \label{lemma}
\end{lemma}
{\bf Proof:}
It is sufficient to show that for any $p$ on ${\cal C}[0]$ outside of the black hole region, there exist sufficiently large $u_0$ and $t_0$, such that
there exists a null geodesic starting from $p$ which does not cross the event horizon and ${\cal C}[t]\cap J^-({\cal I}^+[u])$
 for any large $t>t_0,\ u>u_0$.

$S_{in}[t]$ and $S_{out}[t,u]$ is a closed subset of $S[t]$ by their definition. 
Since they are not intersecting (proposition \ref{prop}, $S_{in}[t]\cap S_{out}[t,u]\neq \emptyset$), suppose $S[t]\backslash S_{in}[t]=S_{out}[t,u]$.  Then $S_{out}[t,u]$ is an open subset of $S[t]$ and $S_{out}[t,u]$ should be a whole of a connected component of $S[t]$ or $\emptyset$. Since $S[t]$ is arc-wise connected, however, from the proposition \ref{prop}, $(S_{in}[t]\cup S_{out}[t,u])\backslash (S_{in}[t]\cap S_{out}[t,u])\neq S[t]$.
Therefore $S[t]\backslash (S_{in}[t]\cup S_{out}[t,u])$ correspond the subset of the null geodesics $N(p,{\cal C}[t])$ and is not empty.
\qed\\

Intuitively, the larger $t$ and $u$ are, the larger number of null geodesics cross $ S_{out}[t,u]$ and $S_{in}[t]$.
 Without any coordinate singularity, for a large time coordinate, a timelike curve has large affine length.
 Therefore null geodesic reaching $S[t]\backslash (S_{in}[t]\cup S_{out}[t,u])$ also has large affine length.
Then our theorem will state that in any long time future, there exists an affinely long null geodesic which does not fall into the black hole and does not escape to the null infinity.

Now we argue the discrimination between the neutral null geodesic and the wandering null geodesic.
In order to describe the situation where a null geodesic is wandering around the black hole we invented the concept of a wandering null geodesic, and therefore, the wandering null geodesic is necessarily a neutral null geodesic.
On the other hand, furthermore, from knowing the structure of the timelike infinity, we might be able to judge the correspondence between them by the global asymptotic structure.
Indeed, Galloway has shown that the existence of a null line without conjugate points suggests the uniqueness of an asymptotically simple spacetime\cite{Galloway2002}.
Nevertheless, it is not our principal aim to investigate that issue here since the structure of the timelike infinity is out of our observational interest.  
Rather, we want to resolve that from the viewpoint of local geometry, especially relating it with the wandering aspects around the black hole.

Since that congruence of neutral null geodesics does not go to infinity, the effect of Riemann curvature must not vanish under the null genericity: $k_{[e}R_{a]bc[d}k_{f]}k^ek^c\neq 0$ anywhere. Then the curvature terms should dominate the deviation equations of null geodesic congruence\cite{Wald:1984rg} in time since expansion $\theta$ and shear $\hat{\sigma}_{ab}$ are dumping by $\sim1/(s-s_0)$, in affine length $s$.
Then $|\hat{\sigma}_{ab}|$ becomes large and the first equation is dominated by the $\hat{\sigma}$-term.
Therefore, soon $\theta$ becomes to take a negative value, which is denoted by $\theta_0$, and then they will have a conjugate point since $\theta$ will diverge within affine length $s\leq 2/|\theta_0|$.

More rigorously, we think of a null generic condition $k_{[e}R_{a]bc[d}k_{f]}k^ek^c\neq 0$.
As the proof of the timelike case in the textbook\cite{Hawking:1973uf}\cite{Wald:1984rg}, the null generic condition and the null energy condition imply the formation of conjugate points, in the proposition 4.4.5 of the textbook\cite{Hawking:1973uf}.
These conditions will hold for physically realistic spacetime since there would be a physical matter and the tangent vector of a null geodesic is not rigorously directed in the principal null directions.

Furthermore, on a complete null geodesic, the generic condition will be satisfied on an infinite number of points.
If the null generic condition and the null energy condition are satisfied, such a conjugate point should appear in time. If so, the number of these conjugate points is not bounded on a complete null geodesic.

\begin{theorem}[commonplaceness of future wandering null geodesics]
Supposing that the spacetime satisfies the null generic condition $k_{[e}R_{a]bc[d}k_{f]}k^ek^c\neq 0$ and the null energy condition $R_{ab}k^ak^b\geq 0$, for any $p$ on ${\cal C}[0]$ outside of the black hole region, there exists a wandering null geodesic starting from $p$.
\label{thm:cpn}
\end{theorem}
{\bf Proof:}
From the lemma \ref{lemma}, for any $p$ on ${\cal C}[0]$ outside of the black hole region, there exists a neutral null geodesic $\lambda_n$ starting from $p$. 
Let $q$ be a point $\lambda_n\cap {\cal C}[t]$. If $t$ is not bounded, $q$ and $p$ are connected by the timelike curve of unbounded length by global hyperbolicity of $M$ (or strong asymptotically predictability). Then the null geodesic can have a parameter according to the parameter $t$ and it gives diffeomorphically an affine parameter.
Since this affine parameter of the null geodesic is unbounded,
the neutral null geodesic can be geodesically complete.
Furthermore, from Ref. \cite{Hawking:1973uf}, if the null energy condition $R_{ab}k^ak^b\geq 0$ is held and the null generic condition
$k_{[e}R_{a]bc[d}k_{f]}k^ek^c\neq 0$ is satisfied in the region generically, there is a conjugate point on the neutral null geodesic.
Besides, since the affine length is infinite, there would be conjugate points infinitely.
\qed\\


By the fact that for all $p$ in a certain region there exists a wandering null geodesic, one might lose the point of discussing the embedding of the set of future wandering null geodesics into the spacetime, because the embedding of it always should cover the causal future of the initial time slicing outside the black hole region (e.g., see figure \ref{fig:subset}).
In this sense, we had better pay attention to the totally wandering null geodesic or the truncated wandering null geodesic\cite{Siino2020}.
Anyway, we will rather concentrate on the local structure of the null geodesic congruence around a future wandering null geodesic in the following section.

Then, as a fundamental issue, we expect that the wandering null geodesic can be regarded as a flag of a black hole spacetimes, in place of the event horizon.
Since the number of the disconnected components of the event horizon can take arbitrary values under the change of timeslicing\cite{Brill2014}\cite{Siino:1997ix}, there is a difficulty in specifying identities of black holes.
On the contrary, since the wandering null geodesic is in a chronal region, its number of disconnected components is free from the ambiguity by the change of timeslicing.
Moreover, to achieve a complete resolution for that problem of identification, we have to take account of the possibility that wandering null geodesics could be present around black hole mimickers (e.g., non-collapsing ultra-compact objects with radius $r$ in the range $2m<r<3m$).
Nevertheless, we are hopeful that it can be excluded by some physical conditions like an energy condition. 
Besides, in the next section we will also give a hint to resolve that by introducing a new concept of `black room'.

Here we want to shortly comment on chaotic aspects of wandering set.
To understand the theorem visually, we had better illustrate the case of the Schwarzschild spacetime.
Then we will find that the catastrophic nature\cite{poston2014catastrophe} of the achronal boundary $\partial J^+(p)$ related to the chaotic appearance of the set of null geodesics near the photon sphere, which include the totally wandering null geodesics.

\begin{figure}[hbtp]
  \begin{center}
\includegraphics[height=10cm]{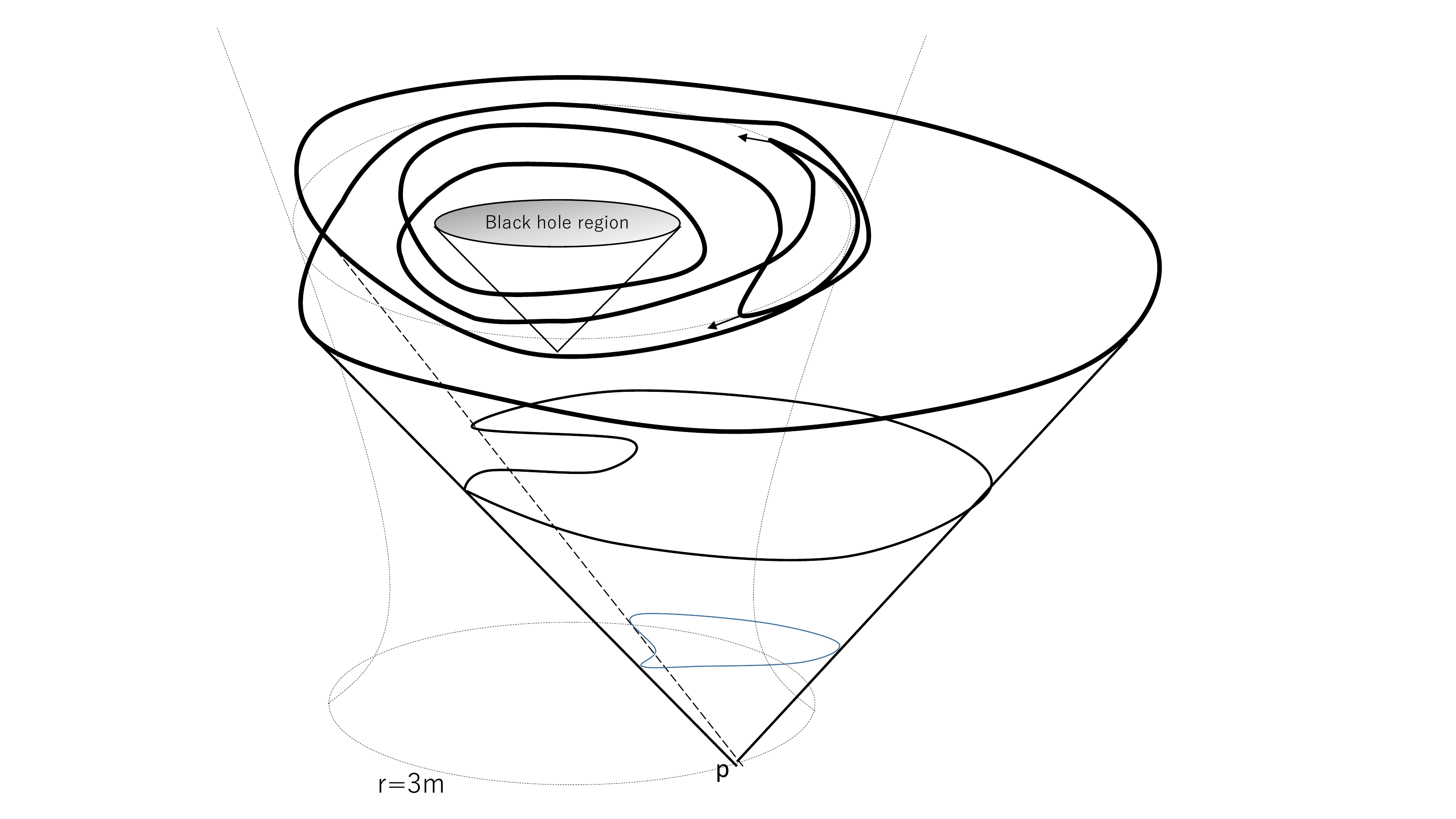}
  \end{center}
 \caption{In the Schwarzschild spacetime, the photon sphere exists on $r=3m$. A null geodesic from the point on the photon sphere will go to inside and outside of 
the sphere $r=3m$, or be still on this photon sphere. They make the wavefront of the null geodesics topologically complicated, after sufficient time has been gone by.  
}
   \label{fig-chaos}
 \end{figure}

In Fig.\ref{fig-chaos}, we are drawing a picture of $S[t]$ in Schwarzschild spacetime, where $p$ is to be on a circular orbit ($r=3m$), as an example.
To watch the evolution permanently, the timeslicing ${\cal C}[t]$ is avoiding the singularity like a maximal slicing in Schwarzschild spacetime\cite{Smarr1978}.
Because of its singularity avoidance, the lapse function is very small inside the event horizon while on the outside of the horizon the timeslice approaches $i^+$ as an affinely parametrized timelike geodesic approaches it.
Then inside the event horizon, the speeds of null geodesic is not fast in this coordinate, which means the advances and enlargement of $S[t]$ is slow there.


Considering a null geodesic going on the unstable circular orbit, it will extend $S(t)$ so that a segment of $S(t)$ meets, crosses, and merges with other separated segments of $S(t)$, while the center of the black hole region is not incorporated by $S(t)$.
Then $S(t)$ gets homotopically non trivial embedding after the crossing.
Moreover, a conjugate point of the wandering null geodesic appears and it enters the inside of $J^+(p)$.

Succeedingly the null geodesic will advance along the circular orbit, while in the direction to the center of a black hole the null geodesic stagnates, and in the outward direction away from the black hole extends $S(t)$.
Therefore, while the null geodesic along the circular orbital cycles around the event horizon several times, its wavefront will get several self-crossover and become chaotically complicated.
Though it is sure from the continuity (theorem \ref{thm:ODE}) that the wavefront is continuous immersion of $S^2$, it might seem to become singular. That will also be a reflection of the catastrophic structure of null geodesics accumulating the wandering null geodesic.

 Considering small deviation from spherical symmetry, however, the chaotic nature may change drastically. 
On the other hand, the catastrophic structure\cite{poston2014catastrophe} will be maintained as stated in the beginning of \cite{Siino2020} by the structural stability.

\section{relation between wandering null geodesics and the black hole shadow}

In our previous work\cite{Siino2020}, we simply discussed the influence of the wandering null geodesic on the brightness of the edge of the black hole shadow, though its statement was something weakened.
Here, however, we will discuss the relation between the wandering null geodesics and the observation of black hole shadow
(If we are careful about the handling of the word shadow, for convenience, we use it to express the difference of bright region and dark region, while it would originally mean only the dark region.).

In the previous discussion, we did not insist that every wandering null geodesic should be relevant for the composition of the photon sphere.
This means that the relevance essentially relies on the accumulation of null geodesics around the wandering null geodesics, and will be corresponding to the concept of stability around the photon sphere in static spherical black hole spacetime.

To find the accumulation in a global sense, the previous discussion might be almost our best, and then further investigation will require any concrete numerical analysis about various null geodesics in order to go beyond the static spherical black hole spacetime.
In the former of this section, our task is to rephrase the discussion with several amendments incorporating the commonplaceness of the wandering null geodesics, and the existence of an event horizon.
That would also answer the question whether we are really looking at the region nearest to the event horizon by black hole shadow observations or not.

In the latter of this section, we will introduce a convenient causal concept named `black room' for describing the geometry and observation of black hole shadow.
With the aid of it, we suggest the optical black hole observation would be related to the wandering null geodesics, more tightly.

\subsection{photon sphere and accumulation of wandering null geodesics}
Now we rephrase the discussion in \cite{Siino2020} for the relevance of the wandering null geodesics to the black hole shadow.

As well known, the black hole shadow in Schwarzschild spacetime is explained by the photon sphere which is the sphere determined by its unstable circular photon orbits. 
Since we have seen that the totally wandering null geodesic coincides with the circular photon orbit in Schwarzschild spacetime, we might expect, in general, the wandering null geodesics are relevant to the black hole shadow.

Here we should be careful about the fact that the existence of an unstable circular photon orbit is relevant to the black hole shadow in the static spherical black hole spacetime.
There, the unstable circular orbit can collect photons and form a photon sphere.

Is the accumulation of the wandering null geodesics relevant to the photon sphere in general situation?
In our previous work\cite{Siino2020}, we demonstrated that the existence of wandering null geodesics should be significantly related to the composition of a photon sphere in the field of view for optical black hole observation.
There, though it may be doubtful that in dynamical situation the black hole geometry can form such a collection of photons which is relevant for the optical observation of the black hole because of the dynamics of the circular orbit, we suggested that from the global viewpoint, this unstable aspect corresponds to the accumulation of the null geodesics in a null geodesic congruence around the wandering null geodesic.

To discuss the structure of the black hole shadow, we would consider the null geodesics related to the event horizon and the wandering null geodesic. 
Then in the sense of global causal structure, we will argue that the accumulation actually occurs and will show that the existence and the configuration of such a past wandering null geodesic mean the existence of at least one past wandering null geodesic which is adjacent to a normal null geodesic. In an asymptotically flat spacetime, it is always the case that such a wandering null geodesic is followed by accumulating null geodesics.


Now we remind readers of the fact that the totally wandering null geodesic and the future (past) wandering null geodesic possibly cause accumulation of the null geodesics, as a result of the commonplaceness of the wandering null geodesics.
The relevance of the set of wandering null geodesics would be a subject to be investigated globally taking account of asymptotic structures of the black hole spacetime.
In the context of global differential structure, such a notion will be realized by the existence of conjugate points.
We suppose a situation where there is at least one past wandering null geodesic.

For an observer $p$ away from the strong gravitation, we will consider a null geodesic congruence along a past wandering null geodesic, whose existence is guaranteed by the similar argument for the commonplaceness of theorem \ref{thm:cpn} in an eternal black hole spacetime.
Then the field of sight is given by a section of the null geodesic congruence before the first conjugate point of the past wandering null geodesic.

\begin{figure}[hbtp]
\begin{center}
 \includegraphics[height=10cm,clip]{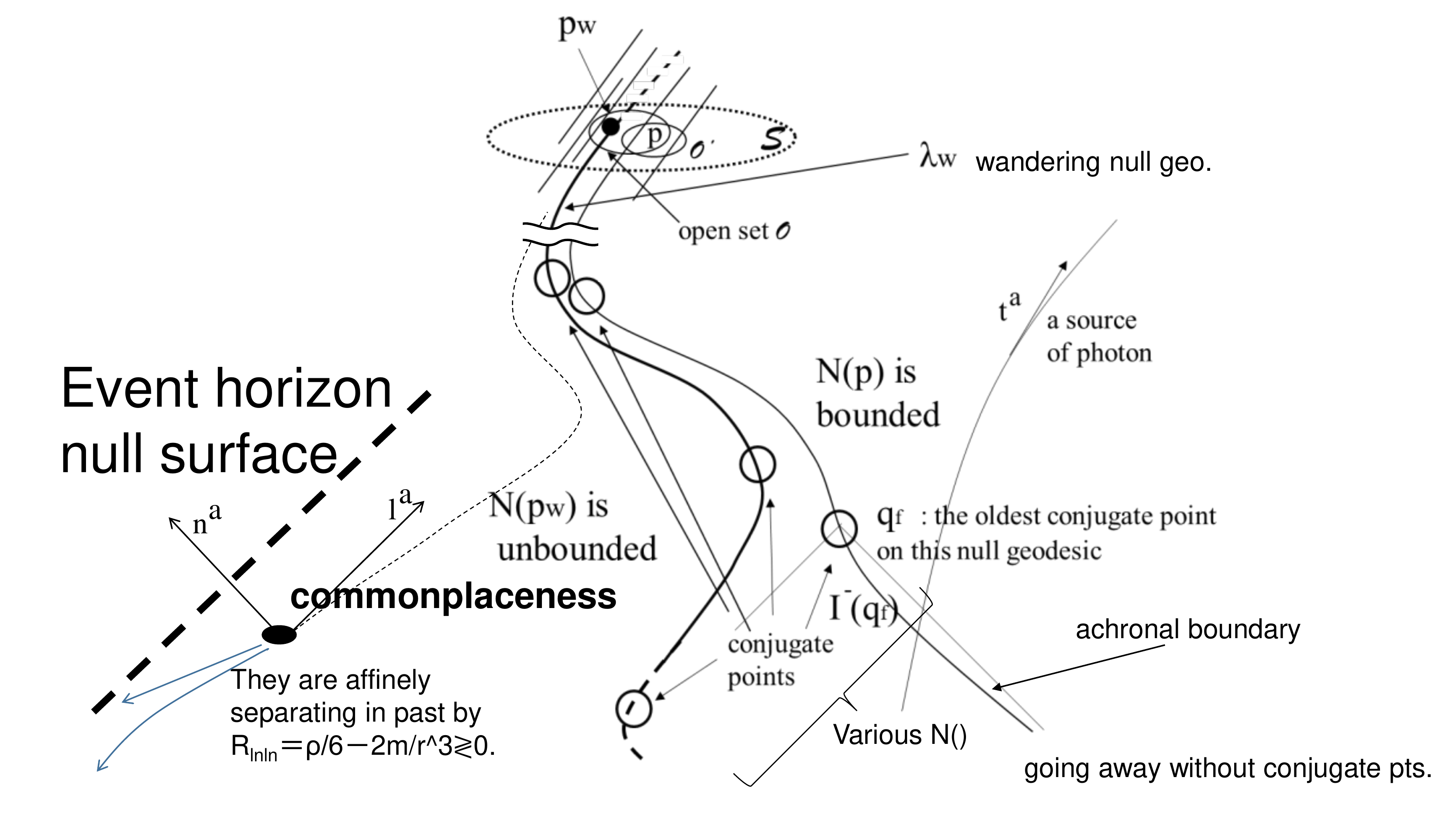}
 \end{center}
 \caption{Singular congruences are considered around a past wandering null geodesic.
 While there are an infinite number of conjugate points on the past wandering null geodesic, there are a finite number of conjugate points on a past directed null geodesic starting from $p$. 
The final (oldest) conjugate point $q_f$ determines a past set $I^-(q_f)$ and a segment of the null geodesic from $q_f$ should be a generator of the boundary of $I^-(q_f)$.
In the star, the sectional curvature is positive and the null geodesic congruence would expand in past direction. That means we cannot receive the information of the spacetime point very close to the event horizon.} \label{fig:5}
\end{figure}

Now we are relaxing the concept by extending the geodesic further beyond the singularity of the Raychaudhuri equation.
 Then the composition of this `singular' congruence is a set of past null geodesics intersecting the field of sight $\cal S$ of the congruence (see Fig. \ref{fig:5}). We denote the congruence with section $\cal S$ as $c[{\cal S}]$.
 The deviation equation will be analytically continued through the conjugate points without a trouble since it is a second order ordinary differential equation along one past null geodesic.
 
 For convenience of discussion, we artificially consider a number function $N(p)$ on $\cal S$ of the conjugate points along the null geodesic congruence.
Suppose that there is a past directed null geodesic congruence along a past wandering null geodesic $\lambda_w$ intersecting $\cal S$ at $p_w$ and the spacetime is geodesically complete in past direction
\footnote{Of course, this is not the case for our cosmology with initial singularity. For such a big bang universe, we will carry out a quantitative investigation, for example the comparison of the magnitude of $N(p)$ with the age of the universe should be significantly discussed.}.
Each geodesic of the congruence $c[{\cal S}]:={\cal S}\times (-\infty,0]$ gives the tangent null vector field $k^a$ of each past directed null geodesics and the past directed exponential map $q_p(t)=\exp (k^a t)\circ p, \ \  (p\in {\cal S}, t\in  (-\infty,0])$ (cf.  the embedding of the congruence into the spacetime is not diffeomorphic since there would be conjugate points).
Then we can give a smooth number function of conjugate points $n (p, t) $ such that 
\begin{align}
&n(p,t):c[{\cal S}]\mapsto \mathbb{R}, \\
&n(p,t_j(p))=j, \\
& t_{i-1}(p)>t>t_i(p)\ \Rightarrow i-1< n(p,t)< i,
\end{align}
where $i$-th conjugate point $q_i[p]$ of $\lambda_p$ is related to the parameter $t_i$ as
$q_i[p]=q_p(t_i)=\exp(t_i k^a)\circ p$. Then, we also consider $N(p)=\max_t\{n(p,t)\}$ for each null geodesic.

Once such a smooth function $n(p, t)$ is found on each $\lambda_p$, from the smoothness of the spacetime and the Stone-Weierstrass theorem\cite{Stone1937}, $N(p)$ is a smooth function in a domain not containing any passing point $p_w'$ of a past wandering null geodesic\footnote{Even if two null geodesic with different number of conjugate points is near at a moment, they can be divided into former past region since they will be a generator of different achronal boundary.}. Note that at $p_w'$, $N(p)$ diverges as $N(p_w')=\infty$.
 
 Suppose an open set $\cal O$ containing $p_w$ includes its open subset $\cal O'$ from which no null past geodesic becomes a wandering null geodesic.
 A segment from $q_f$ of such a past directed null geodesic, where $q_f$ is the oldest conjugate point (the final one in the past direction), should be the generator of $\partial I^-(q_f)$ under the global hyperbolicity (as illustrated in Fig.\ref{fig:5}).

On the other hand, this boundary $\partial I^-(q_f)$ is the achronal boundary which is a closed embedded achronal $C^{1-}$ submanifold\cite{Hawking:1973uf}.
We consider a Cauchy surface $\cal C$ and a smooth timelike vector field $t^a$ whose integral curve can be regarded as a model for orbits of source objects of photons, since every integral curve of $t^a$ intersects $\cal C$ precisely once. 
Then homeomorphism $\psi: \partial I^-(q_f)\mapsto {\cal C}$ is given by the integral curves.  Therefore, the image of $\partial I^-(q_f)$ assigned by ${\cal A}=\psi[\partial I^-(q_f)]$ is homeomorphic to the $C^{1-}$ submanifold $\partial I^-(q_f)$. After all, $\cal A$ is a closed and open subset in the topology of $\cal C$, so that $\cal A$ is $\cal C$.
Then all the orbits of the source intersecting the Cauchy surface $\cal C$ crosses this achronal boundary precisely once.

The `sub-congruence' $c[{\cal O'}]$ is also complete and inextendible, and $\psi[c[{\cal O'}]]$ does not have measure zero in $\cal C$.  
Then, with a non-compact Cauchy surface on which there is no bias in the distribution of luminous sources, each null geodesic of the sub-congruence $c[{\cal O'}]$ will cross an infinite (numerous) number of orbits for the luminous sources since the past directed null geodesic is complete and inextendible---though these may be far from astrophysical generality.

Since $N(p'_w)$ diverges and $N(p)$ will be smooth on the domain ${\cal O}\backslash\{ p'_w\}$, the range of $N(p)$ is containing various positive integers, and is extremely wide there.
Different positive integers of $N(p)$ mean that null geodesics had orbited the black hole for different periods of time before arriving at the observer. 
Consequently, in the analogy with the circular orbit, the null geodesics orbiting the black hole many times are accumulated there. 

Furthermore, as the solution to the deviation equation is smooth, the interval between two conjugate points lying along a null geodesic is finite.
Therefore, since the null geodesics would be orbiting the black hole, past null geodesics departing from the cycling orbit during one cycle, will generally go out to the different directions.
Then they will be destined for entirely different regions at far in the past. Therefore, Theorem \ref{thm:cpn} suggests that every specetime point can emit photons to the neighborhood of any wandering null geodesic, and thus, the black hole collects the light signals from the whole of the universe.
So, the closer $\cal O'$ approaches $p_w$, the wider the range of $N(p)$ becomes.

On the contrary, for a congruence $c[{\cal O''}]$ of a close $\cal O''$ from which every null geodesic is the past wandering null geodesic, at least a class of smooth timelike vector fields ${t'}^a$ exists such that $\psi[c[{\cal O''}]]$ is a compact subset of $\cal C$, which imply there is a strong contrast between $\cal O'$ and $\cal O''$ on the field of view concerning the brightness/darkness around the past wandering null set. 
Of course, such a designed choice of $t^a$ may significantly affect the result.
 

Thus, the accumulation of the null geodesics contributes to the existence of the photon sphere.
Nevertheless, we cannot expect the fade-out of the image near the outside of the wandering null geodesics, as the accumulation implies also the dissipation of the null geodesics because of the time-symmetricity of the concept.
 Indeed, if the spacetime is static and the source of photon is uniformly distributed on the celestial sphere, the brightness of the image outside of the shadow must be uniform.
Probably, however, this accumulation of the null geodesics, coupled with the distribution of the source of light around the black hole would be incorporated to the discussion in order to understand the ring structure of EHT \cite{Akiyama:2019cqa,Akiyama:2019brx,Akiyama:2019sww,Akiyama:2019bqs,Akiyama:2019fyp,Akiyama:2019eap}.

These ideal discussions logically indicate that past wandering null geodesics arriving at the observer are relevant in an optical observation.
For example, that would be corresponding to the fact that, as well known in static spherically symmetric spacetime, the relevance of the photon sphere to the black hole shadow depends on the stability of the null geodesics around the circular orbit of a photon rather than on the aspect of a single photon running on the same circular orbit.

 From the asymptotic flatness, for a sufficiently large $\cal S$ it is natural to assume that not all null geodesics of the congruence $c[{\cal S}] $ can be wandering null geodesics. Then, there must be a wandering null geodesic $\lambda_{rw}$ from $p_{rw}$ such that any small $\cal O$ containing $p_{rw}$ must include an open subset $\cal O'$ not containing any passing point of a past wandering null geodesic.
 Therefore, there should exist at least one such a past wandering null geodesic $\lambda_{rw}$ relevant to the black hole shadow.
  In the case of a spherically symmetric spacetime, this result can be translated as the following.
 Whether a circular orbit of a photon is unstable or not is decided by whether an extremal value of the effective potential of null geodesics is maximal or not.
If there is an extremum in the potential function, it must be a local maximum since the potential is downwardly convex at a large radius by the asymptotic flatness.
Since the location of the maximum corresponds to an unstable circular orbit, there exists a past wandering null geodesic that is relevant to the black hole shadow observed in the asymptotically flat spacetime.
  
As one would easily see, in order to discuss the dark areas of the photon image in detail, it is not sufficient to be concerned with the presence of the event horizon as well as the wandering null geodesic.
To treat the dark areas of the photon image, we would apply the commonplaceness of the wandering null geodesic. 
Theorem \ref{thm:cpn} states that from any starting point we have a future wandering null geodesic in a black hole spacetime.
The starting point corresponds to the spacetime point where a photon source radiates the light signal.
Then, in principle, all of the light source in the world can contribute to the bright regions in the present black hole shadow.

On the other hand, we can also consider the contribution of the event horizon.
Are we really looking at the region close to the event horizon? We cannot simply conclude that the event horizon is black by the black image of the black hole shadow, since the line of sight in the direction of the shadow never reaches to the event horizon in backward time by definition of it.

Nevertheless, from the commonplaceness theorem, there is also a future wandering null geodesic from each spacetime point near but outside the event horizon.
This means that a photon from the direction of the dark region that corresponds to just an inner side of the wandering set seems to approach the event horizon in past direction. Moreover, for a further inner region, which is the near center of an image, we can simply regard the interior of the images of the wandering set should be approximately from an event horizon if the black hole is eternal.
Of course, since the spatial configuration of the light sources around the black hole and the configuration of null geodesics near the event horizon may not be so simple as the case of the Schwarzschild black hole, we cannot illustrate the general images without concrete analysis of null geodesics.

In these arguments, however, the assumption of the eternal black hole is essential and too severe. In a realistic situation, we will require to incorporate the age of the black hole.
If the black hole is dynamically formed like an Oppenheimer-Snyder spacetime, the spacetime point around the event horizon will have two types of contribution to make the shadow image, since in this case there can be light sources near the event horizon unlike the stationary black hole.
The type of contribution is depending on the geometry of the null geodesic congruence, which is controlled by the sectional curvature $R_{abcd}n^al^bn^cl^d$.
By the geodesic deviation equations, that means whether a photon with a four-wave vector $l^a$ propagating in the neighborhood of the event horizon gets away from it in the sense of the affine length in the direction of $n^a$.

To make discussion simple, we suppose the collapsing star is transparent and consider the null geodesic congruence around a future wandering null geodesic starting from one point-source of photons near the event horizon.
The sectional curvature includes both components of Ricci curvature $\sim R/6+R_{nl}$ and the Weyl curvature $C_{nlnl}$.
Since the dominant one among these two quantities behaves as $\rho/6$ in a homogeneous isotropic dust star\cite{Oppenheimer:1939ue}, and $-2m/r^3$ in vacuum Schwarzschild spacetime, respectively,
it will be positive inside of the star where the tidal force is not strong, and negative outside of the star where the tidal force dominates.
Then the null geodesic under consideration does not approach the event horizon in the past direction of $l^a$. 
Consequently, we cannot get the information from the point near a horizon segment in the star, in principle.

Furthermore, in the case of the formation of the black hole, if the age of the dynamical black hole is sufficiently large also we may be able to observe the information infinitesimally close to the event horizon from the image of a black region of the black hole shadow.
On the other hand, if the black hole is young, we might observe the direct image of the background of the star in the center direction of the shadow, since this line of sight will go back to the formation of black hole and can pass through the center of the collapsing star.
\subsection{black room and wandering null geodesics}
Here we want to reveal the role of the wandering null geodesic against the dark region of the black hole shadow.
In the discussion in the Schwarzschild spacetime, the essence of the black hole shadow has been explained mainly by two aspects of photon orbit.
One is that many photon orbits accumulate onto the unstable circular orbit. 
And then the photon sphere becomes bright\cite{Keir2016,Cardoso2014,Cunha2017a}.
In the previous subsection, we have discussed this mechanism in general situations by studying the wandering null geodesics as a generalization of the unstable circular orbit. The other is the fact that there is no orbit escaping from the inner side of the photon sphere after entering it in a Schwarzschild spacetime.
From now on, we develop a corresponding discussion in general situations, with a help of the characteristics of the wandering null geodesics investigated above.
First of all, we introduce auxiliary concept accompanying the black hole, which we define as a new subset of the spacetime manifold.

We suppose that a region named `black room' $\cal R$ has the following property: Any photon that enters the black room from outside region cannot go out as long as it propagates along a null geodesic. Note that differently from the fact that any light signal cannot go out from the black hole, we allow the possibility that a photon generated inside the black room might go out.

The black hole region is an extreme example of the black room, though it does not possess its own exit.
Then it will be possible that several different black rooms exist in one black hole spacetime. If there is the maximal one, it will play the universal role in the optical observation of the black hole, (though one, of course, cannot help being disturbed by the particular configuration of the light sources).

\begin{definition}
Suppose each arc-wise connected (spatially) bounded open subset, ${\cal R}_i\subset M, (i=1,2...)$ such that no null geodesic can escape from ${\cal R}_i$ after entering ${\cal R}_i$, exists.
If there exists ${\cal R}_{max}\in \{ {\cal R}_i| i=1,2...\}$ which includes all ${\cal R}_i$'s, we call ${\cal R}_{max}$ maximal black room.
\end{definition}
Obviously, if there exists a black room, light sources outside $\cal R$ cannot contribute to the image of $\cal R$, so that $\cal R$ becomes dark in optical imaging without a light source contained by $\cal R$. 
This is just analogous to the case of the photon sphere of Schwarzschild spacetime\footnote{The existence of the black hole seems to be essential for the black room. The photon sphere not accompanied by a black hole might not mean the existence of a black room.}.
In Schwarzschild spacetime, there exists the maximal black room whose outer boundary corresponds to the photon sphere. 
And then photons never come from the direction of the maximal black room, provided that there is no source of photons in the maximal black room. 
In realistic situation, it is unlikely that a source of photons is located around the photon sphere or the maximal black room. The light signal originated from realistic light sources, e.g. the accretion disk, would have reached us after accumulating on the boundary of the maximal black room.

Up to now, we have considered the formation occurs at some time. In such a situation, the black room is expected to form as well. However, since the formation of the black room mean cases of eternal black holes. We now consider the situation where the black hole s that all null geodesics escaping from the region had entered the region before, the allowed black room is just the black hole region\footnote{ 
To avoid such an argument related to the black room formation, we think that it is required to consider any local and instantaneous definition of the black room like the apparent horizon indicating the existence of the event horizon.}.
Then we cannot expect the black room to play a remarkable role in all general situations of black hole observation. Knowing that, however, we will suggest the relation between the maximal black room and the wandering null geodesic. 
We want to study the composition of the boundary of the maximal black room.
The null generator of an event horizon or a wandering null geodesic could be lying on the boundary $\partial \cal R$. Nevertheless, we will not strongly concern with the horizon generator since it should not be relevant to the optical observation of the black hole by definition.

From the global reason, we should consider a spacetime region chronologically complete in future direction $\sim\{{\cal C}(t)| t\in (0,\infty)\}$.

\begin{proposition}
A maximal black room $\cal R$ is bounded by a non-spacelike hypersurface which is including at least one null geodesic possibly admitting endpoints at each point\footnote{Though, for a null surface, the null geodesic is in only one direction, there are possible $S^1$-direction of angle for a timelike hypersurface.}.
Under the null energy condition and null generic condition, the null geodesics are the wandering null geodesic, horizon generator, or null geodesic with future endpoint. 
\label{prop:room}
\end{proposition}
{\bf Proof:}
Firstly, we think of $\partial {\cal R}$ which is smooth around $p\in \partial {\cal R}$.
The light cone at $p$ defines all possible null directions on $\partial {\cal R}$ and then corresponding null geodesics $l_p$'s are determined.
Then there can be a small neighborhood ${\cal U}_p \ni p$, such that $\partial {\cal R}\cap{\cal U}_p$ is smooth in its own null directions, whose null  geodesics $l_p$'s has no crossover with $\partial \cal R$ except for $p$, and no conjugate point through $p$.

If $l_p$ is not contained in $\partial\cal R$, we have three possible types of $l_p$ on ${\cal U}_p$, i.e., ($\alpha$) $l_p$ is not in $\rm int\cal R$, ($\beta$) $l_p$ is in $\cal R$'s closure $\overline{\cal R}$, and ($\gamma$) $l_p$ crosses $\partial \cal R$ while it is tangential to $\partial \cal R$ at $p$.

For the type ($\alpha$) where $l_p$ is not in ${\rm int}\cal R$, we easily see that slightly shifted null geodesic $l_p^\prime$ which is almost parallel to $l_p$  have two intersections with $\partial \cal R$ so that it is escaping from $\cal R$ after entering $\cal R$. Therefore, once there is such a null geodesic $l_p$ of type ($\alpha$), the considering black room is immediately denied.
For the third type ($\gamma$), the considering black room also is immediately denied by considering another point $p^{\prime\prime}$ in this direction on $\partial \cal R$ very close to $p$ in which $l_{p{\prime\prime}}$ is of the type ($\alpha$) from the fundamental fact of the Morse theory\footnote{The Morse functions form an open, dense subset of all smooth functions in the $C^2$-topology.}. Then $l_p$ is on $\partial \cal R$ in one side of $p$ and enters $\cal R$ through $p$. Therefore, this type of $l_p$ will be discussed  as the non-trivial endpoint of the null geodesic in the same sense of indifferentiable situation as stated below.

Since the possible directions of $l_p$ are at most $S^1$, we should consider the case (A) where all null geodesics are of the type ($\alpha$), the case (B) where all null geodesics are of the type ($\beta$), and the case (C) where both null geodesics the type ($\alpha$) and the type ($\beta$) exist, while the case (A) has already been excluded by the above argument about the type ($\alpha$) null geodesics.
In the case (B) where all $l_p$'s contained in $\overline{\cal R}$, we see that there can be a point $p^{\prime\prime}$ outside of $\overline{\cal R}$ which gives a null geodesic $l_{p^{\prime\prime}}$ in each null direction through $p^{\prime\prime}$ so close to $l_p$ in the same null direction in the topology Eq.(\ref{top}) that $l_{p^{\prime\prime}}$ has two intersections with $\partial\cal R$ in both sides of $p''$ on ${\cal U}_p$. 
And then, there exists small deformation ${\cal R}^\prime\supset \cal R$ of $\cal R$ in ${\cal U}_p$ such that $\cal R'$ has all $l_{p^{\prime\prime}}\cap {\cal R}^c$'s on its boundary $\partial \cal R'$.
Of course, ${\cal U}_p$ can include no null geodesic escaping $\cal R^\prime$ after entering $\cal R^\prime$.
There is such a small ${\cal U}_p$ that $l_{p^{\prime\prime}}$ never enters $\cal R$ even outside of ${\cal U}_p$ since $\cal R$ is open subset.
Then this case is not allowed, in order to avoid giving another black room which is not subset of $\cal R$.
In the case (C), while not all $l_p$'s are of the type ($\beta$), not any $l_p$ is of the type ($\alpha$). 
Then $l_p$ should be included in $\partial\cal R$ in several directions.
Therefore, at least one null geodesic $l_p$ is on $\partial\cal R$ around the smooth points $p$ on $\partial\cal R$\footnote{Denoting the extrinsic curvature of $\partial {\cal R}$ as $K_{ab}$, the situation is expressed as $K_{ab}k^ak^b \leq 0$ for arbitrary tangential null vectors $k^a$, and at least one photon must satisfy the equality.
The equality will hold only when all the possible null geodesics are included in the surface where we might regard it as a kind of photon surface\cite{Claudel:2000yi}.
}.

If the null geodesic is not a wandering null geodesic but a future complete null geodesic, after the final conjugate point $p_c$ it can be the null geodesic generating a null surface $\partial J^+(p_c)$ (We call such a null geodesic merely a null generator from $p_c$.).
If the null geodesic $l_p$ includes a null generator, the null generator should be a horizon generator from the bounded nature of the definition for the black room.
Then a null geodesic lying on $\partial{\cal R}$ should be a future incomplete null geodesic, future wandering null geodesic or a null geodesic including horizon generator.
Nevertheless, global hyperbolicity (or strong asymptotically predictability) results that the incomplete null geodesic will enter the event horizon provided that it is inentendible. Therefore, since an event horizon is also a black room, we only accept geodesically incomplete null geodesics with a future endpoint such that $\cal R$ cannot be the maximal one.


Next, we discuss the smoothness of the boundary $\partial {\cal R}$.
After knowing the null geodesics around the each smooth points on $\partial {\cal R}$, we will see their tangent future null vector field $k^a$ almost everywhere continuous. 
Then the null vector field should have a critical point at the point where $\partial {\cal R}$ is indifferentiable.
In the context of the Morse theory, at a trivial critical point $p$ with $index_p(k^a)=0$, though $S^1$ of $k^a$-direction does not always coincide with $S^1$ of $-k^a$-direction without the smoothness, they are homeomorphic at this trivial critical point.
So, it is convenient to consider that this situation is an acute limit of the above smooth three cases (A), (B), and (C). 
Then, the allowed null-indifferentiable surface is the limit of the case (C) where in several null directions $l_p$ is on the boundary $\cal R$ and in other directions the boundary is concave and indifferentiable.

On the contrary, at the non-trivial critical point $p$ with $ index_p(k^a)\neq 0$ the change of the topology on the spatial section arises. 
For the non-trivial critical point, however, if $\partial {\cal R}$ is not null surface around $p$ but timelike hypersurface, we easily see that slightly shifted null geodesic from the indifferentiable point have two intersections with $\partial \cal R$ so that it is escaping from $\cal R$ after entering $\cal R$. 
Therefore, such non-trivial endpoints of null geodesics on a timelike hypersurface of $\partial \cal R$ is not allowed.
Then only a null generator provides such endpoints concerning the topology change of the maximal black room.

Consequently, we have seen that a null geodesic around $p$ on $\partial\cal R$ is included in the boundary $\cal R$ in at least one null direction.

\qed\\

Here we briefly comment about the future endpoints of null geodesics lying on the boundary of $\cal R$. 
As shown above, the endpoint of the null geodesic lying on the boundary of the maximal black room is only allowed to be that of a null generator. 
Then the surface generally can become indifferentiable hypersurface.
It seems that the future endpoint of the null generators related to the remarkable dynamical process of the black hole, that is a change of black hole topology\cite{Siino:1997ix}\cite{Siino:1998dq}, black hole evaporation, and so on.
Nevertheless, it will be pretty difficult to find any universal view to restrict the future endpoints. For example, though the future endpoint may be related to the evaporation of black holes, we could not conclude anything in general by the lack of knowledge about the black hole evaporation. 
In the following subsection, however, we attempt to discuss the simple case of coalescing binary black holes, as an example concerned with only past endpoints.

Finally, we want to argue the general possibility of the maximal black room.
Incidentally, one may think that the maximal black room does not generally exist.
Of course, in the Schwarzschild spacetime, there should be the maximal black room, which is bounded by the circular photon orbit corresponding to a totally wandering null geodesic.
What would be the observational result of the absence of the maximal one? 
In general situation, one might think the dynamics or anisotropy prevent the maximal one, (e.g., supposing vibrating or rotating black room).
From the viewpoint of the observation, we may be able to translate this aspect as unsharpeness or deformation of the black shadow images\cite{Wielgus2020}.

\subsection{discussion: binary black hole}
 
In addition to above things, we are interested in the black hole binary since it is a typical source of the gravitational radiation\cite{Abbott:2016blz}.
Here we will also discuss the topological notion of the black hole shadow\cite{Okabayashi2020,Cunha2018}.

Intuitively, the commonplaceness of the wandering null geodesic may imply there would be the families of the wandering null geodesics related to either original black hole or the final black hole, though that is not so convincing because of a chaotic nature of the null geodesics around the wandering null geodesic as illustrated in the bottom of the third section. For clearness of discussion, we will assume that the original two black holes already exist even in the initial state of the universe.

Firstly, we consider all of the wandering null geodesics in this system.
In the considering region of spacetime $M\sim \{{\cal C}(t), (t\in (-\infty,\infty))\}$, there are several causally homotopic classes of the null geodesics\cite{Friedman1993}.

In our standpoint, however, the binary black hole is not defined in mathematical argument of causal structure, but by astrophysical argument which is strongly related to the concept of an evolution of a single star, since we cannot discriminate the merging event horizon and the prolate event horizon under the general covariance of coordinate transformation\cite{SIINO2011}.
It has been argued that rather wandering null geodesics might determine the definite mathematical concept for the topology of the event horizon, e.g. in the case of the binary of black holes\cite{Siino:1997ix,Siino:1998dq,Husa:1999nm,Cohen:2011cf,Bohn:2016soe,Emparan:2017vyp}.



Here, we shall consider the topological notion of the maximal black room for this system, because initially the black holes existed.
If we have a maximal black room for this system fortunately, we would be able to select the wandering null geodesics dominating the black hole shadow observation.
If there is always a maximal black room in a black hole spacetime, at least the concept of the topology of the shadow will become clearer than before since the boundary of the black room is almost timelike in many typical cases.
One would investigate the topology of the maximal black room or topology of its spatial section, even in the case like a binary coalescence.
It seems that the past endpoint of the wandering null geodesic which is lying on the boundary of the maximal black room might be related to the formation or coalescence of it.
Then the indifferentiability of the shadow might be discovered, related to the black hole shadow observation for the arising black hole. 
In other words, a `crease set for black room' may be able to determine most of the structure, incorporating wandering nature of null geodesics as well as the crease set of the event horizon determines the topology changing of it\cite{SIINO2011}\cite{Siino2002}.

We now assume that the maximal black room $\cal R$ is bounded by null geodesics without future endpoints.
And then, if coalescence of binary black holes happens, some of such null geodesics should have past endpoints for the topology changing.
Nevertheless, the past endpoint of the wandering null geodesic cannot be the endpoint for such a merging $\partial {\cal R}$ (in the sense of bounding null geodesics).
If so, the past endpoint must be a crossover point of two timelike hypersurface, but this is not consistent to that two surfaces are bounding the maximal black room, since near the endpoint there is a null geodesic escaping from $\cal R$ after entering $\cal R$ (proposition \ref{prop:room}).
Therefore, the endpoint should be those of the horizon generators.
If the merger finally settles down to a Schwarzschild black hole, however, the measure of these event horizon generators against the photon sphere would be convergent to zero, where in the sense of present study the final state may be concave and indifferentiable and the indifferentiable points are on the null generator.
That will explain the result of some preceding studies\cite{Cunha2018}\cite{Okabayashi2020} suggesting that two black hole shadows will not merge. 




\section{summary}

To clarify the role of wandering null geodesics in optical observation of the spacetime region around a black hole, we have investigated the commonplaceness of the wandering null geodesics and black room region in a black hole spacetime.
Though the wandering null geodesic emanates from an arbitrary spacetime point commonly, it is noticed that the dark rays does not always come from the spacetime points close to the surface `event horizon' in collapsing black hole spacetime, and we cannot conclude that the event horizon is black by watching the black region of the black hole shadow.
Rather, the darkness of the shadow will be the result of the existence of the black room in which there is no escaping null geodesic after entering it.
Under the certain conditions, we have found the maximal black room which dominates universally the dark region in an optical observation of a black hole is almost bounded by the set of wandering null geodesics.

\begin{acknowledgments}
 This work has included the relation to the earlier discussion with Daisuke Yoshida and Haruki Terauchi.
\end{acknowledgments}

 \bibliographystyle{unsrt}
\bibliography{cosmopolitan4} 

\end{document}